\documentclass[aps,prd,onecolumn,showkeys,floats,showpacs, nofootinbib]{revtex4}
\usepackage{graphics,epsfig,color} 
\usepackage{amsmath,amssymb}

\begin{document}

\title[2PN/RM gauge invariance in Brans-Dicke-like scalar-tensor theories]{2PN/RM gauge invariance in Brans-Dicke-like scalar-tensor theories}

\author{Olivier Minazzoli}

\affiliation{Jet Propulsion Laboratory, California Institute of Technology,
4800 Oak Grove Drive, Pasadena, CA 91109-0899, USA}
\begin{abstract}
In this note we study the 2PN/RM gauge invariance structure of a \textit{Brans-Dicke-like} Scalar-Tensor Theories (STT) without potential. Since the spherical isotropic metric plays an important role in the literature, its 2PN/RM STT version is deduced from the general equations given in \cite{moiCQG11}, by using the invariance structure properties. It is found that the second order Eddington parameter $\epsilon$ can be written in terms of the usual post-Newtonian parameter $\gamma$ and $\beta$ as $ \epsilon=4/3 \gamma^2 + 4/3 \beta - 1/6 \gamma  -3/2$.

\end{abstract}

\pacs{04.50.Kd,04.25.Nx}
\maketitle
\section{Introduction}

Because of their expected accuracies, most projects relying on precise measurements of the characteristics of the propagation of light in the solar-system require the knowledge of the space-time metric at the $c^{-4}$ level \footnote{The $c^{-4}$ metric is also called the 2PN/RM metric; where PN/RM stands for Post Newtonian/Relativistic Motion. It means that the development of the Post-Newtonian metric is developed to the order that has to be taken into account when dealing with test particles with relativistic velocities only \cite{moiCQG11,moiPRD09}.} in order to have sufficiently accurate equations describing the various observables (see for instance \cite{klionerCQG10,moiCQG11,dengPRD12} and references therein).

In a recent work \cite{moiCQG11} give the 2PN/RM metric of Scalar-Tensor Theories (STT) without potential, in a set of coordinates that respect the Weak Spatial Isotropy Condition (WSIC: $g_{ij} \propto \delta_{ij}+O(c^{-4})$, \cite{DSX91}); otherwise arbitrary. Although eventually such kind of dynamical metric would have to be taken into account for accurate solar system calculations, a lot of works exploring the 2PN/RM phenomenology simplify the problem by considering the spherical case in isotropic coordinates \cite{epsteinPRD80,fischbachPRD80,richterPRD83,turyshevCQG04,plowmanCQG06,ashbyCQG10,teyssandierARXIV12}. This allows to explore the main features of the 2PN/RM phenomenology; without the complication of a more realistic dynamical metric as given for instance by \cite{moiCQG11}. Hence, a parameterized 2PN/RM metric is often considered, with a new parameter $\epsilon$ entering in the $c^{-4}$ space-space part of the metric.

In this note, we first study the gauge invariance structure left in the 2PN/RM field equations, which are such that the coordinate system respects the WSIC only. Then we use this analysis in order to derive the value of the second order post-Newtonian parameter $\epsilon$ in the case of STT without potential. This last calculation illustrates the method that uses the gauge invariant structure of the field equations in order to find specific gauges.
\section{Field equations}

We start with the usual STT action without potential in the Jordan representation \cite{moiCQG11}:
\begin{eqnarray}
S=\frac{c^{4}}{16\pi G}\int d^{4}x\sqrt{-g}\left[ \Phi R-\frac{\omega \left(
\Phi \right) }{\Phi }g^{\alpha \beta }\partial _{\alpha }\Phi \partial
_{\beta }\Phi \right] +\int d^{4}x\sqrt{-g}L_{NG}\left( \Psi ,g_{\mu \nu
}\right) .\nonumber\\  \label{jrepr}
\end{eqnarray}
$g$ is the metric determinant, $R$ is the Ricci scalar constructed from the \textit{physical} metric $g_{\mu \nu}$ \footnote{In our model, we assume that $g_{\mu \nu}$ is the \textit{physical} metric, in the sense that it is the one that describes actual time and length as measured by clocks and rods in our experiments \cite{GEF2004}.}, $\mathcal{L}_m$ is the material Lagrangian, $\Psi$ represents the non-gravitational fields. From there, one deduces the following 2PN/RM metric as well as the corresponding field equations \cite{moiCQG11}:
\begin{eqnarray}
g_{00} &=&-1+\frac{2W}{c^{2}}-\beta \frac{2W^{2}}{c^{4}}+O\left(
c^{-5}\right),  \label{jmetric} \\
g_{0i} &=&-\left( \gamma +1\right) \frac{2W_{i}}{c^{3}}+O\left( c^{-5}\right),
\nonumber \\
g_{ij} &=&\delta _{ij}\left\{ 1+\gamma \frac{2W}{c^{2}}+\left( \gamma
^{2}+\beta -1\right) \frac{2W^{2}}{c^{4}}\right\}+\left( \gamma +1\right) 
\frac{2W_{ij}}{c^{4}}+O\left( c^{-5}\right) , \nonumber
\end{eqnarray}
with
\begin{eqnarray*}
\gamma \equiv \frac{\omega _{0}+1}{\omega _{0}+2}, ~~\beta \equiv 1+\frac{\omega _{0}^{\prime }}{\left( 2\omega _{0}+3\right) \left( 2\omega
_{0}+4\right) ^{2}},~~ G_{eff}\equiv \frac{2\omega _{0}+4}{2\omega_{0}+3}G,
\end{eqnarray*}
and
\begin{eqnarray}
&&\square W+\frac{1+2\beta -3\gamma }{c^{2}}W\triangle W+\frac{2}{c^{2}}\left(
1+\gamma \right) \partial _{t}J =-4\pi G_{eff}\Sigma +O\left( c^{-3}\right),\nonumber \\
&&\triangle W_{i}-\partial _{i}J =-4\pi G_{eff}\Sigma ^{i}+O\left(
c^{-2}\right),  \nonumber \\
&&\triangle W_{ij}+\partial _{i}W\partial _{j}W+2\left( 1-\beta \right) \delta
_{ij}W\triangle W-\partial _{i}J_{j}-\partial _{j}J_{i}-2\gamma \delta
_{ij}\partial _{t}J=-4\pi G_{eff}\Sigma ^{ij}+O\left( c^{-1}\right), 
\nonumber \\
\label{fieldeqWij} 
\end{eqnarray}
Where one has set 
\begin{eqnarray}
J &=&\partial _{t}W+\partial _{k}W_{k}+O\left( c^{-2}\right), \nonumber \\
J_{i} &=& \partial _{k}W_{ik}-\frac{1}{2}\partial _{i}W_{kk}+\partial _{t}W_{i}-%
\frac{1-\gamma }{2}\partial _{i}P, \label{eq:ji}
\end{eqnarray}
with
\begin{equation}
\triangle P+2\frac{\beta -1}{1-\gamma }W\triangle W-2\partial _{t}J  =-4\pi G_{eff}\frac{\Sigma ^{kk}}{3\gamma -1}+O\left( c^{-1}\right), \label{fieldeq2} 
\end{equation}
while for the matter part of the equations, one has:
\begin{eqnarray*}
&&\Sigma =\frac{1}{c^{2}}\left( T^{00}+\gamma T^{kk}\right), ~~~\Sigma ^{i} =\frac{1}{c}T^{0i}, ~~~ \Sigma ^{ij} =T^{ij}-\gamma T^{kk}\delta _{ij}.
\end{eqnarray*}

\section{The gauge invariance structure}

As in the General Relativity (GR) case, the diffeomorphism invariance leaves 4 degrees of freedom (dof.) left unconstrained in the field equations. Therefore, there is a \textit{gauge-invariance-like} behavior of such field equations. Because of the WSIC ($g_{ij} \propto \delta_{ij}+O(c^{-4})$) imposed to the metric at the 1PN level, the 1PN field equation gauge freedom is characterized by an arbitrary scalar function $\lambda$ only. Indeed, the WSIC in the Jordan representation follows from the Strong Spatial Isotropy Condition (SSIC: $g_{00} g_{ij}=-\delta_{ij}+O(c^{-4})$, \cite{DSX91}) in the Einstein representation \cite{moiCQG11}. Therefore, it fixes the gauge freedom corresponding to the spatial dof. that would appear in the field equations otherwise; and leaves only the gauge freedom corresponding the choice of time coordinates. One has:
\begin{eqnarray}
\label{GIW}
W'=W-\frac{1}{c^2} \partial_t \lambda, ~~~~W'_i=W_i+\frac{1}{2(1+\gamma)} \partial_i \lambda.
\end{eqnarray}
At the 2PN/RM level, the metric in general can't be put in an isotropic form anymore, and the spatial dof. reappear anew in the field equations, as in GR \cite{moiPRD09}. Therefore, at this level, the gauge invariance is characterized by an additional arbitrary 3-vector $A_i$, such that:
\begin{equation}
W'_{kl}=W_{kl} + \partial_k A_l + \partial_l A_k + \frac{\gamma}{1+\gamma} \delta_{kl}~ \partial_t \lambda. \label{eq:tranfoij}
\end{equation}
But to be complete, one also has to take into account the invariance of the equation on the scalar field $P$:
\begin{eqnarray}
\label{GIP}
P'=P+\frac{1}{1+\gamma} \partial_t \lambda.
\end{eqnarray}
This scalar field is a \textit{leftover} of the scalar field $\Phi$ in the field equations at the $c^{-4}$ level. It is due to the fact that the scalar field equation's source is proportional to the trace of the stress-energy tensor instead of being proportional to $\Sigma$.\\
Equations (\ref{GIW}-\ref{GIP}) represent the 2PN/RM gauge invariance structure of the scalar-tensor field equations in a set of coordinates that respect the WSIC.

\section{The spherical isotropic case}

The spherical isotropic case has an important place in the literature related to the 2PN/RM metric. Indeed, it looks like the simplest metric one can use in order to derive the 2PN/RM phenomenology characterized for instance by the time transfer or deviation angle equations. In this case -- corresponding to a spherical source at the center of the coordinates -- the metric in various alternative theories would write, according to \cite{epsteinPRD80}:
\begin{eqnarray}
g_{00} &=&-1+\frac{2W'}{c^{2}}-\beta \frac{2W'^{2}}{c^{4}}+O\left(
c^{-5}\right),  \label{eq:metriced1} \\
g_{0i} &=&O\left( c^{-5}\right), \\
g_{ij} &=&\delta _{ij}\left\{ 1+\gamma \frac{2W'}{c^{2}}+\frac{3 \epsilon}{2} \frac{ W'^{2}}{c^{4}}\right\}+O\left(
c^{-5}\right) , \label{eq:metriced3} 
\end{eqnarray}
with $W'=GM/r'$, where $G$ is the effective gravitational constant, $M$ the mass of the source and $r'$ the radial coordinate in the isotropic system of coordinates. We dub $\epsilon$: second order Eddington parameter -- equal to 1 in GR. However, one should notice that there is no reason to expect that a vector-tensor theory for instance, would not break the spherical symmetry of the problem because of the local direction of the space part of the vector field. Therefore, such a metric seems to be useful for a very restricted set of alternative theories -- namely, probably only scalar-tensor theories. However, while in our opinion one should use the metric in the general form, such as the one given by equations (\ref{jmetric}), it still seems interesting to give the value of the $\epsilon$ parameter in scalar-tensor theories -- mainly because most of the papers studying some aspects of the 2PN/RM phenomenology use the metric in the form given by equations (\ref{eq:metriced1}-\ref{eq:metriced3}) \cite{epsteinPRD80,fischbachPRD80,richterPRD83,turyshevCQG04,plowmanCQG06,teyssandierIAUS10,ashbyCQG10,teyssandierARXIV12}.\\
 In order to get the metric in this class of coordinate system, one has to find a gauge transformation that kills the anisotropic terms in equation (\ref{fieldeqWij}). This can be achieved by realizing that $\partial_i W \partial_j W =\frac{1}{8} \partial^2_{ij} W^2 + \delta_{ij} U$, where $W=GM/r$ -- $r$ being the original radial coordinate (ie. in no specific coordinate system) -- and $U=\frac{1}{4} (GM/r^2)^2$. 
Therefore, since we are considering a static vacuum field solution where $\triangle W=O(c^{-2})$, $\partial_t J=0$ and $S_{ij}=0$, equation (\ref{fieldeqWij}) writes: 
\begin{equation}
\triangle W_{ij}+\frac{1}{8} \partial^2_{ij} W^2+\delta_{ij} U-\partial _{i}J_{j}-\partial _{j}J_{i}=O\left( c^{-1}\right), \label{eq:Wijsphere}
\end{equation}
which can be re-written as:
\begin{equation}
\triangle W'_{ij}+\partial_i \left(\triangle A_j+\frac{1}{16} \partial_j W^2 -J_j  \right)+ \partial_j \left(\triangle A_i+\frac{1}{16} \partial_i W^2 -J_i  \right)=-\delta_{ij} U+O\left( c^{-1}\right),
\end{equation}
by using the gauge invariance of the field equation as well as the commutativity of partial derivatives.
Then by choosing the 3-vector gauge field $A_i$ that satisfies the following equation:
 \begin{equation}
\triangle A_i= J_i - \frac{1}{16} \partial_i W^2= \partial_k W_{ki}- \frac{1}{2} \partial_i W_{kk} +\partial_t W_i - \frac{1-\gamma}{2} \partial_i P - \frac{1}{16} \partial_i W^2,\label{eq:transfA}
 \end{equation}
 \footnote{On the contrary, if one wants to write the metric in the harmonic gauge \cite{moiCQG11}, one needs to impose: $\triangle A_i= J_i $ as in General Relativity \cite{moiPRD09}.} -- which is obviously invertible -- equation (\ref{eq:Wijsphere}) re-writes:
 \begin{equation}
 \triangle W'_{ij}=- \delta_{ij} ~\frac{1}{4} \left(\frac{GM}{r'^2}\right)^2+O(c^{-1}) \rightarrow W'_{ij}=- \delta_{ij}~ \frac{1}{8} W'^2 +O(c^{-1}),
 \end{equation}
with $r'=r+O(c^{-2})$. From there, after injecting into the space-space component of the metric (\ref{jmetric}), one gets:
 \begin{equation}
 \epsilon = \frac{4}{3} \gamma^2+\frac{4}{3} \beta -\frac{\gamma}{6} -\frac{3}{2}.
 \end{equation}
It is the result found in \cite{damourPRD96} after they considered directly the 2PN spherical body solution in isotropic coordinates of the STT field equations (see their equation (5.8)). As a corollary, setting $\gamma=1$ and $\beta=1$ in (\ref{eq:transfA}) gives the corresponding transformation in the GR case.

\begin{acknowledgments}
This research was supported by an appointment to the NASA Postdoctoral Program at the Jet Propulsion Laboratory, California Institute of Technology, administered by Oak Ridge Associated Universities through a contract with NASA. \copyright 2012 California Institute of Technology. Government sponsorship acknowledged.
\end{acknowledgments}
\bibliographystyle{jphysicsB}


\end{document}